\documentclass[a4paper]{revtex4}
\usepackage{epsfig}
\usepackage{hyperref}
\begin{document}
\title{In-plane oxygens in high-temperature superconducting cuprates}
\author{G.~Nik\v{s}i\'{c}}
\author{I.~Kup\v{c}i\'{c}}
\author{D.~K.~Sunko}
\email{dks@phy.hr}
\author{S.~Bari\v{s}i\'{c}}
\affiliation{Department of Physics, Faculty of Science, University of
Zagreb,\\ Bijeni\v cka cesta 32, HR-10000 Zagreb, Croatia.}
\begin{abstract}

The role of the oxygen degree of freedom in the cuprates' superconducting
planes is analyzed in detail. Structural and photoemission results are
reviewed to show that the most sparse description of the in-plane electronic
states requires explicit control of the oxygens. For metallic states, the
relative contributions of oxygen and copper vary along the Fermi surface (FS),
with the arc metallicity dominantly oxygen-derived. For the magnetic
responses, we find that the observed incommensurability arises naturally if
one keeps separate the roles of the two sites. For the charge order in LBCO,
we propose a scenario, based on magnetic interactions in the plane. We stress
the need for further experimental investigations of the evolution of the
intracell charge distribution with doping, and for a better theoretical
understanding of the large particle-hole-symmetry breaking required for
successful phenomenologies, but difficult to reconcile with ab initio
calculations.

\vskip 1cm\noindent
\textbf{Key words:} Emery model, magnetic fluctuations, electron-doped
superconductors.

\noindent
\textbf{Running title:} Oxygens in high-T$_c$ cuprates

\end{abstract}

\maketitle

\section{Introduction}\label{intro}

High-temperature superconductors have so far evaded attempts to reduce them to
a simple model expression of the relevant physics. The physical reason for
this is that they lie in a number of crossover regimes, between Mott
antiferromagnetism (AF), different manifestations of metallicity, and
superconductivity (SC), the precise nature of which is still a matter of
considerable controversy. Simplicity usually corresponds to a limiting model
situation, rather than to an interpolation between different limits, in
particular the covalent and the ionic, as seems to be the case with the
cuprates.

In the present work, we hope to convince the reader that the key simplifying
step to a physical understanding of the cuprates is to take into account two
physically distinct, and equally important, degrees of freedom in the
copper-oxide planes, namely the coppers and the oxygens (Emery model). We find
that further reduction to a one-band model is counterproductive, because it
ends by reintroducing the particle-hole ($ph$) symmetry breaking, an essential
feature of the experimental picture, in a physically less transparent way than
if the oxygens were retained in the first place.

Investigating the relative role of oxygens and coppers in both the
conductivity and SC presents a number of experimental opportunities, to
distinguish the highest-T$_c$ cuprates from cuprates with a lower T$_c$, or
from similar materials, which are not superconducting. In our view, high-T$_c$
SC cannot be reduced to its essentials in general, without first understanding
what makes the concrete cuprates so special.

\section{Critical experiments}\label{smoke}

\subsection{The LTT tilt}

In the 214 class of high-T$_c$ compounds, Ba-doped LCO and Nd-doped LSCO
simultaneously develop a low-temperature tetragonal (LTT) phase, and a
precipitous drop in T$_c$, around 1/8 hole
doping~\cite{Moodenbaugh88,Tranquada95}. Thus the LTT tilt appears to be a
critical probe into the high-T$_c$ mechanism. It specifically affects the
oxygens in the SC planes, splitting the degeneracy of $O_x$ and $O_y$ sites
within a unit cell. A static intracell charge transfer is therefore expected
to appear with the tilt, which may stabilize it, via the analogue of the
Peierls mechanism~\cite{Barisic90}.

The tilts are fairly small, about $3^\circ$, and have a large RMS scatter,
nearly twice the mean value~\cite{Haskel00}. While this observation led some
to doubt~\cite{Haskel00} the coherent Peierls-like picture, it can only
emphasize the critical sensitivity of the SC to a perturbation of the oxygen
degeneracy. Significantly, the conductivity remains metallic, indeed with
strong SC fluctuations above T$_c$ around 1/8 doping~\cite{Tranquada08}. Hence
in this case SC does not vanish because an insulator has appeared, but because
oxygen degeneracy is critical to SC coherence in high-T$_c$ cuprates, for
reasons remaining to be discovered.

\subsection{Zinc substitution}

The Zn-substitution effect is universal among the cuprates. The same 2--3\% of
Zn in place of Cu in the CuO$_2$ planes destroys SC at optimal doping both in
hole-doped LSCO~\cite{Tarascon87} and in electron-doped
NCCO~\cite{Barlingay90}. No structural change is involved, hence the effect is
a local electronic probe into the SC mechanism. Zn is in the d$^{10}$
configuration, precluding any hole occupancy on the Cu site, which reverts the
neighboring oxygens back into the O$^{2-}$ state of the parent
insulator~\cite{Mazumdar89}. Both the Zn $d$-orbitals and O $p$-orbitals are
closed, so the unit cell acts as a zero boundary condition for the holes on
the CuO$_2$ lattice.

Significantly like the LTT tilt, Zn substitution does not destroy
conductivity, albeit it introduces an finite upturn in the residual
resistivity, similar to those uncovered by magnetic-field suppression.
Changing the ionicity of the oxygens by Zn amounts to a strong perturbation of
the relative oxygen site energy, pointing, again, to the critical importance
of site order on the oxygens for the SC, as distinct from the conductivity, in
the cuprates.

Zinc most effectively destroys SC in underdoped YBCO and low-T$_c$ compounds,
which only show a Fermi arc in ARPES. In optimally doped YBCO, and,
importantly, overdoped compositions, where the FS is well developed even if
T$_c$ is not the highest, it is much less effective~\cite{Fukuzumi96}. By
topological arguments within the Emery model, the wave-functions near the vH
singularities are copper-dominated, while, as we shall see below, the Fermi
arcs are oxygen-dominated in the parameter regimes required for a successful
phenomenology. Hence it would appear that the oxygens are primary in the SC
pairing, while the coppers play a supporting role, most simply imagined as a
density-of-states effect, although other scenarios, involving the mechanism of
the SC pairing around the antinodal point more directly, are also possible.

Nickel is markedly less effective than Zn in destroying SC~\cite{Hudson01}.
The open $d$-orbitals in Ni may hybridize with the oxygens, so the latter
remain in the covalent limit. We thus expect Ni substitution to suppress T$_c$
less efficiently in underdoped cuprates on the hole side, and in all cuprates
on the electron side, because it affects the oxygens less, which we regard as
primarily responsible for the SC in those compounds. In optimally hole-doped
or overdoped cuprates, its effects should increase with the role of Cu sites
in SC.

\section{Covalency in the three-band picture}

\subsection{Hopping overlaps}

It has long been recognized that the three-band model has to be applied to
ARPES in the cuprates in a physical regime, different than the one inferred
from quantum-chemical calculations~\cite{Qimiao90}. The latter would have a
Cu--O overlap $t_{pd}\gg t_{pp}$, the O$_x$--O$_y$ overlap, which however
results in a FS, rotated by 45$^\circ$ with respect to experiment. It was
noticed very early~\cite{Kotliar88} that the large on-site repulsion $U_d$ on
the coppers effectively reduces the value of $t_{pd}$. However, that effect
cannot explain the dominance of $t_{pp}$ in ARPES.

First, the physical regime predicted by \emph{ab initio} approaches is borne
out by high-energy XPS measurements~\cite{Veenendaal94}: XPS and ARPES really
observe different high- and low-energy physical regimes, respectively. More
revealingly, to fit ARPES, it is not sufficient to decrease $t_{pd}$: one must
have a $t_{pp}$ absolutely larger than any \emph{ab initio} calculation to
date has justified. The measured dispersion of the open band in the
$(0,0)$--$(\pi,\pi)$ (diagonal) direction is about $0.8$~eV wider below the
Fermi level (i.e. $\sim 1.5$~eV overall) than can be justified \emph{ab
initio}~\cite{Meevasana07}. Adding magnetic correlations is expected to
decrease $t_{pp}$, hence the discrepancy is not merely large, but goes the
wrong way, as well. Because $t_{pp}$ is uniquely dispersive in the diagonal
direction, by topology of the lattice alone, copper-based correlation effects
are poor candidates to explain the absolute scales of the open bands in BSCCO
and YBCO, although they can depress the ratio $|t_{pd}/t_{pp}|$.

Finally, there is no indication of a $t_{pd}$-driven paramagnetic band-width
collapse near the metal-insulator transition, as expected in the mean-field
framework~\cite{Brinkman70}. Chemical potential data are consistent with the
picture that the first doped holes already occupy a hybridized
band~\cite{Ikeda10}. Simply, coherent hopping remains possible even in the
presence of a large (dynamic) charge and spin disorder on the coppers,
indicating, again, the importance of $t_{pp}$. The insulating state at small
dopings is reached in this picture via superexchange, which orders the coppers
at the expense of the residual oxygen conductivity.

\subsection{Site energies}

The site energies show a similar discrepancy between high-energy (XPS) and
low-energy (ARPES) experiments. While XPS confirms the LDA prediction of a
large charge-transfer gap $\Delta_{pd}$, i.e. $U_d>\Delta_{pd}>|t_{pd}|>
|t_{pp}|$, ARPES fits require quite a different regime, $4|t_{pp}|\,
\raisebox{-1mm}{$\displaystyle\stackrel{\displaystyle >}{\sim}$}
\,\Delta_{pd}> t_{pd}^2/\Delta_{pd}$. Clearly the band parameters must be
strongly renormalized to pass from one to the other. The physical distinction
of the latter is that it brings the bonding ``copper" band in anti-crossing
with the next, ``oxygen" band (non-bonding at $t_{pp}=0$), such that the Fermi
arcs of the hybridized open band are in fact oxygen-like. This is most
striking in NCCO~\cite{Sunko07}. The metal-insulator transition itself has
been proposed to proceed by delocalization of oxygen holes following an
O$^{2-}$$\to$O$^{1-}$ orbital transition~\cite{Mazumdar89}. The low-energy
regime, seen in ARPES, is the normal state precursor to SC. It is in this
low-energy regime that SC is critically sensitive to perturbations of the
oxygen sites, as discussed above.

It is still an open question, to what extent the regime, in which the
(renormalized) copper and oxygen states anticross close to the Fermi level, is
common to the cuprates. The rapid evolution of the FS with doping in LSCO
suggests, assuming the Hartree-Fock (HF) picture, a similar evolution of the
Cu vs.\ O charge content in the SC compositions. On the other hand, notable
early XAS~\cite{Bianconi87} and EELS~\cite{Nucker88} experiment indicates
nearly complete oxygen dominance of the conducting bands in both LSCO and
YBCO, while the latter shows little FS evolution with doping. Evidently the
site content in the conducting states may not be directly inferred from the HF
three-band model fits. A more sophisticated accounting is currently under way,
based on the idea~\cite{OSBarisic12}, that a significant redistribution of
copper spectral strength into localized states is possible when the Hubbard
$U_d\to\infty$, because the charge-transfer scale then appears as the dominant
``slow'' counterpart to the ``fast'' interference of the empty upper and
filled lower Hubbard band, the latter pushed to infinity at infinite $U_d$,
with finite spectral weight. The mechanism by which this occurs is a kinematic
waiting effect, namely an itinerant hole cannot hop on the copper site if
another is already there, but has to wait for the other to depart. This
waiting also gives rise to an effective scattering $U_{d\mu}\sim
t_{pd}^4/\Delta_{d\mu}^3$ between itinerant states on oxygens, where,
critically, $\Delta_{d\mu}$ is the difference between the energy of the d$^9$
copper site level and the chemical potential, not the oxygen level as in the
usual expression for superexchange. Hence a magnetic scale appears near
optimal doping, where superexchange effects are presumably small, by virtue of
the oxygen metallicity.

\subsection{Relation to the one-band model}

To fit experiment, additional hopping terms are usually added to the $t$--$J$
model, the first of which, $t'$, breaks $ph$ symmetry, as does the $t_{pp}$
term in the three-band model. Often even further terms are introduced, turning
the model dispersion into an effective Fourier-Taylor series, which can easily
fit any observed FS~\cite{Eschrig03}. However, when $ph$-symmetry-breaking
terms are properly disentangled~\cite{Sunko09} from the symmetry-preserving
ones, it turns out the correction outweighs the initial term by a factor of
3--8 in realistic fits, because the relevant ratio is $|4t'/(t+2t')|$, with
$t'/t<0$, not $\sim|t'/t|$, as one might naively assume.

Such dominance of the correction means that the mapping of the three-band onto
the one-band dispersion is singular in the physical regime actually observed
in the cuprates, as also indicated by unrealistically long-distance hopping
terms, amounting to a diverging Taylor series. Reduction to the one-band
dispersion thus involves considerable difficulty in establishing the
physical meaning of comparisons with experiment, and in addition control over
the relative spectral weight of Cu and O sites is lost. Retaining both coppers
and oxygens explicitly is in our view the appropriate simplification, allowing
for their different physical roles.

\section{Magnetic responses}

Magnetic response in the SC cuprates is generally
incommensurate~\cite{Reznik04,Dunsiger08,Enoki12}. We have previously
connected the peak-rotation and ``arc-protection'' effects~\cite{Niksic12}. 
Here we present a simple qualitative mechanism, by which the complexity of the
cuprates can give rise to an incommensurate magnetic response already at
low order, one-loop level.

First, the magnetic interactions of the hybridized extended states are
mediated by scattering on the coppers, hence the vertices of the $ph$ bubble
involve the corresponding projectors. Because the two legs of the bubble are
at different $k$-vectors, convoluting them can give a minimum at the
commensurate $(\pi,\pi)$ point, if $t_{pp}$ is strong enough. The second
effect is the known weakening of the log-squared SDW response at the vH
filling to log when $ph$ symmetry is broken~\cite{Lederer87}, e.g.:
$\ln^2\omega\rightarrow\ln|2t'/t|\ln\omega$ in the one-band model. Again, if
$t_{pp}$ is large enough, the prefactor can weaken the logarithm
quantitatively, as in $\ln 1\approx 0$. Finally, the logarithm is sensitive to
disorder: even a $1$~K broadening in the single-particle lines is enough to
suppress it below the incommensurate peaks surrounding the minimum created by
the projectors. The situation is summarized in Fig.~\ref{project} for the vH
filling itself, noting that realistic SC compounds have an open large Fermi
surface, hole-doped relative to it.

All three effects are due to copper-oxygen complexity. Site projectors are
relevant in multiband situations, especially if the interaction is localized
in real space. The weakening of the logarithm is due to $ph$ symmetry
breaking, not present in simplified models like $t$--$J$. The qualitative
effect of weak disorder is only expected to be universal in materials
sufficiently complex, for a small broadening to be an intrinsic property. The
ensuing incommensurate AF is in the strong-coupling limit, because the side
peaks are finite, so a transition can be obtained in magnetic mean field only
for a finitely strong spin-flip coupling vertex.

\section{Charge and spin order in LBCO}

In the LTT phase of LBCO, both charge (CO) and spin (SO) collinear stripe
order have been observed in the depression of SC near the 1/8
anomaly~\cite{Hucker11}. The wavelenghts are very short,
$\mathbf{q}_{CO}=0.54\pi/a$ for $x=1/8$, and connected by the interesting
relation $\mathbf{q}_{CO}=2(\mathbf{q}_{SO}-\mathbf{Q}_{AF})$. The transition
temperatures are always arranged $T_{LTT}\ge T_{CO}>T_{SO}$, with
$T_{LTT}=T_{CO}$ for $x<1/8$. The coupling of SO to CO via a thir$d$-order
``S$^2$C'' invariant was proposed previously~\cite{Barisic09}, which accounts
for the factor of 2 in the wave-vectors. The principal remaining question is
whether the driving mechanism comes from CO or SO.

At fillings $x>1/8$, $T_{CO}$ is markedly less than $T_{LTT}$, and
$\mathbf{q}_{CO}$ appears as in a second-order transition, while LTT is
first-order. For $x\approx 1/8$, the two temperatures coincide, while for
$x<1/8$ we concur~\cite{Hucker11} that the apparent coupling, $T_{LTT}=T_{CO}$
for $x<1/8$, is most probably a background effect: the observed CO can only
exist in the LTT phase. The steady lowering of $T_{CO}$ below $T_{LTT}$ for
$x>1/8$ is naturally understood if the two transitions are not due to a
divergence of the same susceptibility, i.e. the driving mechanisms are
different.

A tentative complete scenario rests on the observation~\cite{Hucker11} of
long-, but not infinite-range SDW correlations, for $T_{SO}<T<T_{CO}$, hence
$T_{CO}$ behaves as $T_{MF}$ (mean-field) for SO. Coulomb interactions are 3D,
while magnetic interactions are 2D, mediated by covalency. We propose it is
really SO which is driving CO via the ``S$^2$C'' invariant, only CO can get
stabilized already at the higher $T_{MF}$, because it is helped by the 3D
nature of the Coulomb forces, while the covalency between planes is negligible
by comparison. This scenario is corroborated by the fact~\cite{Hucker11} that
CO melts in all three directions simultaneously. It is also conventional to
expect that short-range Coulomb interactions can contribute to the distinction
between the LTT and LTO backgrounds, although the precise manner remains to be
determined. Conversely, for a phonon to be responsible, a mode would have to
exist, of which we are not aware, with significantly different dynamics in the
two phases.

\section{Discussion}

The metallization of the cuprates against a strong ionic background has been a
persistent challenge both to theory and interpretation of particular
experiments. Our picture is broadly based on the crossover, at extremly small
dopings, between the AF-assisted Mott-localized state and the state, which is
a mixture of coherent extended, and incoherent single-particle states,
localized within a CuO$_2$ unit cell. The extended states are O-dominated,
while the localized ones are Cu-based. Their localization is due to incoherent
Cu--O charge transfer excitations within the CuO$_2$ unit cell, analogous to
the mixed-valence fluctuations in heavy-fermion systems. On the other hand,
the discommensuration of magnetic correlations is associated with spin-flip
excitations between extended states. Recent NMR investigations have directly
been interpreted in terms of a two-component picture~\cite{Haase12}.
Similarly, observations of a $T^2$ dependence of the conductivity in the
pseudogap state~\cite{NBarisic12} indicate a Fermi-liquid-like metallic
component~\cite{Mirzaei12}, which we believe is oxygen-based.

Direct O--O hopping changes the AF at the vH filling from commensurate to
incommensurate. Its effect on the metallic band is even more dramatic, as the
\emph{ab initio} approaches, which start from a copper d$^9$ state, are unable
to account for a large part of the dispersion in the $\Gamma$--$M$ direction.
This may be related to a long-predicted orbital transition with
doping~\cite{Mazumdar89}, from Cu$^{2+}$--O$^{2-}$ to Cu$^{1+}$--O$^{1-}$,
which gives a plausible microscopic interpretation of the oxygen-dominated
anticrossing regime, to which we were independently led by ARPES in NCCO,
although, to repeat, reproducing ARPES requires a large $t_{pp}$ in all
cuprates.

A two-component picture is indicated for SC as well, although it is too early
to pronounce definitively on the relative roles of Cu and O sites. SC seems
first to appear on the oxygens, although it is more robust where coppers are
also involved at the Fermi level. We trust the simple and transparent
two-component picture, based on the physical lattice states, to play an
important role in the eventual understanding of the high-T$_c$ cuprates. We
suggest further experimental effort, in particular with local probes, to be
directed at disentangling the roles of the two sites.

\acknowledgments

This work was supported by the Croatian Government under Project
No.~$119-1191458-0512$.


\newpage

\newcounter{Lcount}
\begin{list}{\textbf{Figure~\arabic{Lcount}.}}
  {\usecounter{Lcount}
  }

\item\label{project}

The static bare susceptibility $\chi_{0}^\pm(\omega=0)$. Thin, thick, and
dashed lines indicate no damping, 1~K, and 10~K damping, respectively. Note
the broad minimum of the background at M $(\pi,\pi)$, due to O--O hopping in
the anticrossing regime. The collinear maxima determine the magnetic
incommensurability. 

\end{list}

\pagestyle{empty}

\newpage
\begin{center}
\includegraphics[height=8cm]{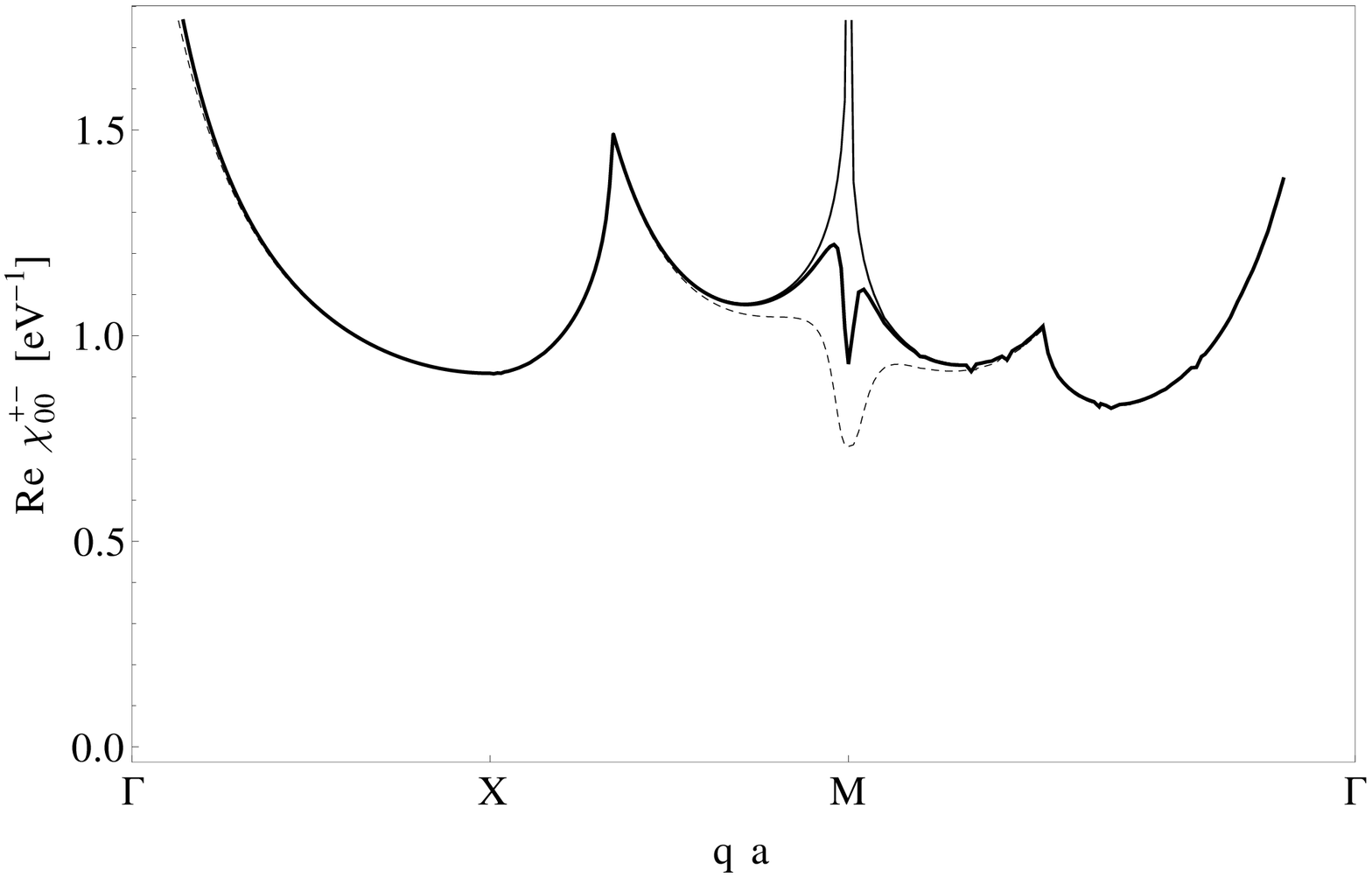}
\vfill
Figure \ref{project}.
\end{center}

\end{document}